# Developing a large scale population screening tool for the assessment of Parkinson's disease using telephone-quality voice

**Siddharth Arora[1, ¶, *], Ladan Baghai-Ravary[2], Athanasios Tsanas[3, ¶]**

[1]Somerville College, University of Oxford, OX2 6HD, UK

[2]Aculab plc, Milton Keynes, MK1 1PT, UK

[3]Usher Institute of Population Health Sciences and Informatics, Medical School, University of Edinburgh, EH16 4UX, UK

[¶]Denotes equal contribution

[*]**Corresponding Author**

Siddharth Arora: Siddharth.Arora@maths.ox.ac.uk

**Contact Information:**

Siddharth Arora, Somerville College, Woodstock Road, University of Oxford, Oxford, UK, OX2 6HD

**Running title:** Assessment of Parkinson's disease using voice

## Abstract

Recent studies have demonstrated that analysis of laboratory-quality voice recordings can be used to accurately differentiate people diagnosed with Parkinson's disease (PD) from healthy controls (HC). These findings could help facilitate the development of remote screening and monitoring tools for PD. In this study, we analyzed 2759 telephone-quality voice recordings from 1483 PD and 15321 recordings from 8300 HC participants. To account for variations in phonetic backgrounds, we acquired data from seven countries. We developed a statistical framework for analyzing voice, whereby we computed 307 dysphonia measures that quantify different properties of voice impairment, such as, breathiness, roughness, monopitch, hoarse voice quality, and exaggerated vocal tremor. We used feature selection algorithms to identify robust parsimonious feature subsets, which were used in combination with a Random Forests (RF) classifier to accurately distinguish PD from HC. The best 10-fold cross-validation performance was obtained using Gram-Schmidt Orthogonalization (GSO) and RF, leading to mean sensitivity of 64.90% (standard deviation, SD 2.90%) and mean specificity of 67.96% (SD 2.90%). This large-scale study is a step forward towards assessing the development of a reliable, cost-effective and practical clinical decision support tool for screening the population at large for PD using telephone-quality voice.

_________________________________________________________________________

# I.    Introduction

Parkinson's disease (PD) is the second most common neurodegenerative disease, and approximately 60,000 people are diagnosed every year in the USA alone; similar incidence rates are reported in Europe (Tanner and Goldman, 1996). Typical characteristic PD symptoms include tremor, rigidity, bradykinesia, and postural instability; critically for this project, voice and speech quality degradation has also been well documented in the PD research literature (Logemann et al., 1978; Harel et al., 2004; Ho et al., 1998; Skodda et al., 2009; Tsanas, 2012; Tsanas et al., 2012; Chen and Watson, 2017). Existing tests for PD assessment and monitoring require the physical presence of the person in the clinic and rely on expensive human expertise. It has been estimated that between 2010 and 2030, the number of Medicare beneficiaries aged over 65 years with PD in the USA will increase by 77% from 300,000 to 530,000 (Dorsey et al., 2013). As the burden of PD is expected to shift from developed western countries to developing eastern countries, remote technologies combined with expert neurologist care could considerably improve the availability and quality of healthcare available to patients. This study proposes investigating novel approaches toward robust, cost-effective, and remote assessment of PD relying solely on voice samples collected over the standard telephone network, hence, facilitating its widespread use as a population screening tool.

Vocal performance degradation is met in the vast majority of people diagnosed with PD, and may be one of the earliest indicators of disease onset (Harel et al., 2004). Using *high-quality* voice recordings, recent studies have developed technologies to: (1) discriminate PD from controls (Little et al., 2009; Das, 2010; Åström and Koker, 2011; Luukka, 2011; Tsanas et al., 2012; Chen et al., 2013; Naranjo et al., 2016; Orozco-Arroyave et al., 2016; Godino-Llorente et al., 2017; Parisi et al., 2019), (2) PD symptom severity telemonitoring (Tsanas et al., 2011; Eskidere et al., 2012), and (3) monitoring voice rehabilitation in PD (Tsanas et al., 2014b). Recent studies have also investigated the feasibility and efficacy of using smartphone technology that extended the use of voice data to include four additional tests for gait, postural sway, dexterity, and reaction times, to support clinical diagnosis for PD. Specifically, using a dataset comprising 10 PD and 10 HC participants, recorded three times daily for a duration of one month using smartphones, an average accuracy of 97% was reported in discriminating PD from HC (Arora et al., 2015). A major limitation of that pilot study, however, was that it was conducted with a very small cohort size.

The aforementioned studies may be limited in scaling massively as a potential screening tool for PD because they rely on expensive specialized equipment to collect the data, typically in a laboratory-based environment. Moreover, these facilities would not be available in resource-constrained settings, thereby limiting their practical use. Also, a vast majority of current studies in the research literature are limited in small sample sizes (<100 participants), and typically only focus on a group of people from the same phonetic background; previous work has emphasized the need to scale-up results in larger cohorts and across multiple phonetic backgrounds (Little et al., 2009; Tsanas et al., 2012).

In this study, we investigate whether telephone-quality voice recordings collected using readily available standard commercial consumer phones could be used to provide easily accessible, cost-effective means towards reliable PD assessment. To the best of our knowledge, this is the largest PD characterization study undertaken using telephone-quality voice recordings.

The manuscript is organized as follows: in section II we present the protocol used for data acquisition along with data summary. In section III, we present the methods focusing on: (A) data preprocessing, (B) feature extraction, (C) feature selection (FS), (D) exploratory statistical analysis, (E) statistical mapping, and, (F) model generalization and validation. Section IV presents the out-of-sample 10-fold cross-validation (CV) results. In section V we summarize the key findings of the study.

## II.  Data

We collected sustained vowel phonations (where participants were prompted to pronounce 'aaah…' for as long and as steadily as possible) through telephone-quality digital audio lines, under realistic, non-controlled conditions. It is also worth noting that dysarthria has been commonly associated with PD, as first suggested by Darley et al. (1969). Interestingly, recent work has also shown that dysarthria can be used to identify participants who are at risk of developing PD, i.e. participants with rapid eye movement sleep behavior disorder (RBD; Rusz et al., 2016). However, given that this study involved collecting recordings from participants from 7 different countries, we decided to focus on analyzing sustained phonations (dysphonia).  The rationale of collecting these vocal sounds lies in the fact that analysing sustained vowel phonations circumvents problems associated with running speech, such as accents and linguistic confounds (Titze, 2000). Moreover, the efficacy of dysphonia analysis for characterizing PD voice has been demonstrated in our previous work (Tsanas et al., 2010; Tsanas et al., 2011; Tsanas, 2012). The telephone-quality voice recordings were collected over a standard digital line as part of the Parkinson's Voice Initiative (PVI)[1]. The vision of PVI was to try and enable radical breakthroughs, through developing voice-based tests as accurate as clinical tests, which can be administered remotely at a low cost. This can be particularly useful for resource constraint settings. The key objectives of PVI are to transform practice by having the following aims: 1) reduce logistical costs associated with diagnosis and monitoring in clinical practice, 2) facilitate high-frequency monitoring that can inform individualized treatment decisions, so as to be able to optimize drug dosage and timing for each individual participants, and, 3) introduce a cost-effective means of mass recruitment of participants for clinical trials.

---

[1] http://www.parkinsonsvoice.org/

To collect the recordings, the project advertised a phone number that participants could call within various countries, and simple verbal instructions were given during the call. Participants provided information regarding their age, gender, and whether a formal clinical diagnosis of PD had been made. An interactive voice response (IVR) system, using the health insurance portability and accountability act (HIPAA)-compliant Aculab cloud, was used to handle the incoming calls. The calls were anonymous and started with a prompt explaining that the call was going to be recorded for research related to PD, and that more information could be found either by pressing a number on the keypad or by going to the PVI website. The callers were told to hang up the call if they did not want to continue and that by continuing the call they were giving consent for the use of their data for research purposes. They were told that they could end the call at any time if they did not want to continue. Callers were also told that only participants aged 18 or more could take part and after asking for their age, any who were under the age of 18 were thanked and the call ended. Participants with essential tremor were eligible to participate and were instructed to answer 'no' when asked if they have been diagnosed with PD. The recordings were not performed in a clinical context. Participants were entirely self-selected. Further details of the study were made available over the phone on request by typing a number on the phone's keypad. All recordings were non-identifiable, and no personal information was stored. The call lasted about three minutes on average. Two sustained phonations from each participant were recorded, sampled at 8 kHz. The data were obtained from the following seven geographical locations: Argentina (144 recordings), Brazil (227 recordings), Canada (1521 recordings), Mexico (75 recordings), Spain (573 recordings), USA (12675 recordings), and UK (4088 recordings). This resulted in a total of 19,303 voice recordings. Summary details about the study were made available to participants by optionally pressing a button when making the call; participants were notified that by continuing the call they would be providing informed consent for their data to be used in this research project.

## III. Methods

### III.A Data Pre-processing

To determine stated diagnostic and other participant data, we developed automated speech recognition of some 60,000 responses to prompts (do you have PD, what is your age, what is your gender, etc.). The automated speech recognition was designed using the mel-frequency cepstral coefficients (MFCCs) (Mermelstein, 1976) extracted from audio prompts and support vector machines (SVMs). If the accuracy to recognize a prompt using the automated speech recognition was less than 90% (as quantified using SVMs), we carefully checked each recording manually to identify the audio prompts. For feature extraction, we ignored recordings for which the prompt was not clear. Inadequate length of the sustained phonation can result in some of the 307 features extracted from the voice recordings in being too noisy. Given the low sampling frequency of recordings (8 kHz), we decided to exclude recordings with phonations less than 2 seconds, as done by previous studies (see Arora et al., 2015;

Arora et al., 2018a; Arora et al., 2018b). After screening out non-usable recordings, we processed 2759 recordings from 1483 PD participants, and 15321 recordings from 8300 control participants. The symptoms (PD/HC) that were self-reported by the participants were treated as the gold standard. Table 1 presents the general characteristics and basic demographics of the participants used in the final analysis. The PD and HC cohorts had a very similar sex ratio (~44% females). On average the PD participants were older than the controls (Table 1). Using the two-sided Kolmogorov-Smirnov test, we could not reject the null hypothesis that the distributions of age for the PD and control participants were realizations from the same underlying distribution (at 5% significance level). To investigate the effect of presbyphonia, we performed additional analysis by quantifying the strength of the relationship between the most discriminatory features with participant age focusing exclusively on the HC cohort. Moreover, in order to investigate any potential effect of gender differences on classification accuracy, we perform analysis for discriminating PD from HC using recordings from: (1) all participants, (2) only female participants, and, (3) only male participants.

## III.B Feature extraction

We use the dysphonia measures that we have been rigorously described in our previous studies (Tsanas et al., 2010; Tsanas et al., 2011; Tsanas, 2012). The rationale, background, and algorithms used to compute these features are explained in detail in those studies. Here, we summarize these algorithms. For convenience, Table 2 lists the extracted features, grouped together into algorithmic "families" of features that share common attributes, along with a brief description of the properties of the speech signals that these algorithms aim to characterize. The articulator features extracted from voice recordings characterize the fluctuations and instability of articulators during sustained vowel phonation (International phonetic alphabet /aː/), and are not used to characterize dysarthria.

A crucial aspect of extracting the dysphonia measures is the computation of the fundamental frequency (F0); its computation is often a prerequisite for the determination of many features. Here, we used the SWIPE algorithm (Camacho and Harris, 2008), which was previously shown to be the most robust and accurate F0 estimation algorithm for sustained vowel /a/ phonations in comprehensive tests using both physiologically plausible data obtained using a sophisticated mathematical model, and also using a database with actual phonations where the ground truth was provided by means of electroglottographic signals (Tsanas et al., 2014a).

Typical examples of features used to characterise sustained phonations are jitter and shimmer (Titze, 2000). The motivation for these features is that the vocal fold vibration pattern is nearly periodic in healthy voices whereas this periodic pattern is considerably disturbed in pathological cases. Therefore, on average we reasonably expect that jitter and shimmer values will be larger in PD participants compared to healthy controls. We remark that there are different definitions of jitter and shimmer, sometimes referred to as jitter variants and shimmer variants (Tsanas et al., 2011), for example by normalizing the dysphonia measure over a range of *vocal fold cycles* (time interval between successive vocal fold collisions). We investigated many variations of these algorithms that we collectively refer to as *jitter* and *shimmer variants* (Tsanas et al., 2011; Tsanas, 2012). The study participants did not include individuals with other Parkinsonian or voice-related disorders. Some of the atypical feature values (such as atypical jitter and shimmer values) which are broadly associated with vocal impairment cannot thus be used as a biomarker of PD across the general population based only on the findings of this study. Building on the concept of irregular vibration of the vocal folds, earlier studies have proposed the *Recurrence Period Density Entropy* (RPDE), the *Pitch Period Entropy* (PPE), the *Glottis Quotient* (GQ), and F0-related measures (Little et al., 2009; Tsanas et al., 2011). RPDE quantifies the *uncertainty* in the estimation of the vocal fold cycle duration using the information theoretic concept of entropy. PPE uses the log-transformed linear prediction residual of the fundamental frequency in order to smooth normal *vibrato* (normal, small, periodic perturbations of the vocal fold cycle durations which are present in both healthy and PD voices), and measures the impaired control of F0 during sustained phonation. GQ attempts to detect vocal fold *cycle durations*. Then, we work directly on the variations of the estimated cycle durations to obtain the GQ measures. The F0-related measures (such as the standard deviation of the F0 estimates) include the difference in the measured F0 with the expected, healthy F0 in the population for age- and gender-matched controls (Titze, 2000). The second general family of dysphonia measures quantifies noise, or produces a *signal to noise ratio* (SNR) estimate.

The physiological motivation for SNR-based measures is that pathological voices exhibit increased aero-acoustic noise, because of the creation of excessive turbulence due to incomplete vocal fold closure. Such measures include the *Harmonic to Noise Ratio* (HNR), *Detrended Fluctuation Analysis* (DFA), *Glottal to Noise Excitation* (GNE), *Vocal Fold Excitation Ratio* (VFER), and *Empirical Mode Decomposition Excitation Ratio* (EMD-ER). GNE and VFER analyze the full frequency range of the signal in bands of 500 Hz (Michaelis et al., 1997; Tsanas et al., 2011).

Additionally, we have created signal to noise ratio measures using energy, nonlinear energy (Teager-Kaiser energy operator) and entropy concepts whereby the frequencies below 2.5 kHz are treated as 'signal', and everything above 2.5 kHz treated as 'noise' (Tsanas et al., 2011). EMD-ER has a similar justification: the Hilbert-Huang transform (Huang et al., 1998) decomposes the original signal into components, where the first components are the high-frequency constituents (in practice equivalent to noise), and the later components constitute useful information (actual signal). Given the limitations of linear modelling approaches in analyzing speech (Little et al., 2006), we used both linear and nonlinear approaches.

Finally, *Mel Frequency Cepstral Coefficients* (MFCC) have long been used in speaker identification and recognition applications, but have shown promise in recent biomedical voice assessments (Godino-Llorente et al., 2006; Fraile et al., 2009; Tsanas et al., 2011; Tsanas et al., 2014b; Orozco-Arroyave et al., 2016). MFCCs do not have a clear physical interpretation regarding the properties of the speech signal they capture: broadly they are aimed at detecting subtle changes in the motion of the articulators (tongue, lips) which can be thought of as complementary additional information to standard vocal fold perturbations (e.g. jitter, shimmer)). Overall, applying the dysphonia measures gave rise to a 18080×307 *feature matrix*.

### III.C Feature Selection

A large number of dysphonia measures available in the study may potentially lead to performance degradation in the statistical mapping phase, a well-known data analysis problem known as the *curse of dimensionality* (Hastie et al., 2009). Although modern state of the art statistical mapping algorithms are, in general, well-versed in alleviating this problem, even powerful classifiers such as RF may suffer in the presence of a high dimensional dataset, and the computational complexity to train the learners is considerable. Therefore, it would be particularly useful if the dimensionality of the original set could be reduced. A reduced feature subset also facilitates inference, facilitating insight into the problem via analysis of the most predictive features (Guyon et al., 2006; Hastie et al., 2009).

An exhaustive search through all possible feature subsets is computationally intractable, a problem that has led to the development of *feature selection algorithms* which offer a rapid, principled approach to reduction of the number of features. FS is a topic of extensive research, and we refer to Guyon et al. (2006) for further details.

Combining a set of different FS algorithms helps overcome the variability associated with a single algorithm (Tsai and Hsiao, 2010). Since each FS technique scores the importance of features based on a unique criterion, the rankings of the most salient features can vary subject to the choice of the algorithm. To account for limitations associated with a single FS algorithm, we used a range of FS algorithms with a very different scoring criterion. Specifically, we employed four efficient FS algorithms: (1) *minimum Redundancy Maximum Relevance* (mRMR) (Peng et al., 2005), (2) *Gram-Schmidt Orthogonalisation* (GSO) (Chen at al., 1989), (3) *RELIEF* (Kira and Rendell, 1992), (4) *Least Absolute Shrinkage and Selection Operator* (LASSO) (Tibshirani, 1996). The mRMR algorithm uses a heuristic criterion to set a trade-off between maximizing *relevance* (association strength of features with the response) and minimizing *redundancy* (association strength between pairs of features). It is a *greedy* algorithm (selecting one feature at a time), which takes into account only pairwise redundancies and neglects *complementarity* (*joint* association of features towards predicting the response). The GSO algorithm projects potentially useful features for selection at each step onto the *null space* of those features that have already been selected in previous steps; the feature that is maximally correlated with the target in that projection is selected next. The procedure is repeated until the number of desired features has been selected. RELIEF is a *feature-weighting algorithm*, which promotes features that contribute to the separation of samples from different classes. It is conceptually related to margin maximization algorithms and has been linked to the k-Nearest-Neighbor classifier (Gilad-Bachrach et al., 2004). Contrary to mRMR, RELIEF uses complementarity as an inherent part of the FS process. Moreover, we use LASSO that has been shown to have *oracle properties* (correctly identifying *all* the 'true' features contributing towards predicting the response) in *sparse* settings when the features are not highly correlated (Donoho, 2006). However, when the features are correlated, some noisy features (not contributing towards predicting the response) may still be selected (Meinshausen and Yu, 2009). All aforementioned FS algorithms have shown promising results over a wide range of different application areas.

The feature subsets were selected using a cross-validation (CV) approach (see Section III.F), using only the training data at each CV iteration, following a voting methodology that we have previously described in Tsanas et al. (2012; 2014b). We repeated the CV process a total of 10 times, where each time the $M$ features ($M = 307$) for each FS algorithm appear in descending order of selection. The feature selection process employed in this study comprised of the following key stages: (1) Balancing: the feature matrix was balanced (by randomly under-sampling observations from the majority class) to ensure equal representation of recordings from PD and controls, (2) Splitting – the balanced feature matrix was split into non-overlapping train and test sets using a 10-fold CV scheme, (3) Selection – the balanced training feature matrix was fed into the four feature selection algorithms specified above (mRMRM, GSO, RELIEF, and LASSO). The above process of balancing, splitting and selection was performed 100 times (10-fold CV with 10 repetitions). The feature indexes which appeared most frequently over the 100 iterations were used to identify the final feature subset for each FS algorithm. The top-ranked features from each algorithm were fed as input into the classifier in the subsequent mapping phase to estimate the binary outcome. We computed discrimination accuracies using a different number of top-ranked features to identify the optimum number of features for each FS algorithm. In addition, we used the majority voting scheme to combine feature rankings from these four FS algorithms to generate a single composite ranking (Tsanas et al., 2012). We refer to the ranking obtained using this majority scheme as *ensemble ranking*.

### III.D Exploratory statistical analysis

In order to gain a preliminary understanding of the statistical properties of the features, we computed the Pearson correlation coefficient and the mutual information $I(\mathbf{x}, \mathbf{y})$, where the vector $\mathbf{x}$ contains the values of a single feature for all phonations, and $\mathbf{y}$ is the associated response. Because the mutual information is not upper bounded, we have followed the strategy to obtain the *normalized* mutual information (Tsanas, 2012): the computed $I(\mathbf{x}, \mathbf{y})$ was divided through with $I(\mathbf{y}, \mathbf{y})$ for presentation purposes in order to ensure that it lies between 0 and 1. The larger the value of the normalized mutual information, the stronger the statistical association between the feature and the response.

### III.E Statistical mapping

The preliminary exploratory statistical analysis in the previous step provides an indication of the strength of association of each feature with the corresponding response. However, ultimately our aim is to develop a functional relationship $f(\mathbf{X}) = \mathbf{y}$, which maps the dysphonia measures $\mathbf{X} = (\mathbf{x}_1 \ldots \mathbf{x}_M)$, to the response $\mathbf{y}$. That is, we need a *binary classifier* that will use the dysphonia measures to discriminate HC from PD participants.

This particular application is a well-studied, highly nonlinear problem where simple linear approaches will likely not generalize well. Hence, we report findings using established state of the art *ensemble learning* statistical machine learning approaches using parallel tree base learners. The topic of ensemble learning has received considerable attention in the research literature because of its potential to map highly nonlinear settings and provide satisfactory outcomes in complex real-world applications (Kuncheva, 2004; Polikar, 2006). Specifically, we used RF which is extremely robust to the choice of its hyper-parameters, and hence we used the default values following Breiman's suggestion (Breiman, 2001), which greatly helped us reduce the computational time associated with the classification. Regarding our choice of classifier, in our previous studies on objective characterization of PD symptoms including voice, we found that the performance of RF to be quite competitive (Arora et al., 2015; Arora et al., 2018a; Arora et al., 2018b; Tsanas et al., 2011; Tsanas et al., 2012; Tsanas et al., 2014b). Moreover, RF have been shown to be fairly insensitive to the choice of two hyper-parameters (Breiman, 2001): (a) the number of trees should be fairly large (due to which we choose 500 trees, which is the default suggestion), and, (b) the number of features considered to construct each branch of a given tree (we chose the square root of the total number of features, which again is the default suggestion). It is due to the aforementioned reasons that we employed RF in this study.

### III.F Model generalization and validation

Validation in this context aims at an estimate of the *generalization* performance of the classification based on the dysphonia features, when presented with novel, previously unseen data. The tacit statistical assumption is that the new, unseen data will have a similar joint distribution to the data used to train the classifier. Most studies achieve this validation using either CV or bootstrap techniques (Bishop, 2007; Hastie et al., 2009). In this study, we used a 10-fold CV scheme, where the original data was split into two subsets: a training subset comprising 90% of the data, and a testing subset comprising 10% of the data. The data was balanced at each CV iteration to account for the difference in PD and HC cohort size. The 10-fold CV process was repeated a total of 10 times, where on each repetition the original dataset was randomly permuted prior to splitting into training and testing subsets. On each repetition, we computed the sensitivity, specificity, accuracy, and balanced accuracy. Errors over the different CV repetitions were averaged, and the process was repeated for a different number of input features (5, 10, 25, 50, 75, 100, 150, 200, 250, 307) selected using four different FS schemes (mRMR, GSO, RELIEF, LASSO) and an ensemble ranking. Figure 1 summarizes the complete methodology.

## IV. Results

The summary measures differed slightly between the two groups (PD and HC). The out-of-sample classification accuracy quantified using two different performance scores (sensitivity and specificity) employing all, female, and male recordings are presented in Figures 2, 3 and 4, respectively. These figures show that the best classification accuracy is obtained using RF-GSO, with overall accuracy figures being in the range of 64-68% (SD~2%). We quantify the binary classification performance for different number of input features, using four FS schemes. The discrimination accuracies obtained using only about 50-100 features were comparable with the accuracy achieved using all 307 features (as evident from Figures 2-4). This is encouraging as a reduced feature subset not only facilitates inference via analysis of the most predictive features, but it also increases the likelihood of any diagnostic support tool developed using this framework to have practical relevance by reducing the associated computational costs.

In addition to using a RF classifier, we used a random classifier (naïve classifier) as a benchmark that we aim to outperform, and this is particularly relevant in cases of unbalanced data. The random classifier benchmark is akin to diagnosing a subject as having PD based on flipping an unbiased coin. Specifically, this classifier assumes $P_{\mathrm{PD}}$:$P_{HC}$ = 0.5:0.5, where $P_{\mathrm{PD}}$ is the probability of a subject being diagnosed as having PD, while $P_{HC}$ is the probability of a subject being identified as a HC. For example, a subject is classified as having PD if the outcome of a fair coin toss is head; else, the subject is identified as a HC. Given that we balance the dataset before training and testing, the chance level for classifying a subject as either PD or control is 50%. The classification accuracy obtained using RF is statistically significantly better than the accuracy obtained using the naïve random classifier, which as expected, resulted in an accuracy of ~50%.

We performed additional analysis to try and gain better insight into vocal impairment in PD by identifying a suboptimal feature subset using high-quality voice recordings. Specifically, we extracted features from 263 voice samples collected from 43 participants (33 PD and 10 controls), whereby the recordings were collected in an industrial acoustics company (IAC) sound-treated booth with a head-mounted microphone (see Tsanas et al. 2012). As opposed to 8Khz recordings used in this PVI study, Tsanas et al (2012) used recordings sampled at 44.1Khz. We selected a subset of 10 highly ranked features from the Tsanas et al (2012) dataset and tested the efficacy of these 10 features on the PVI dataset. Using top 10 salient features from the Tsanas et al (2012) dataset, we achieved a mean balanced accuracy of 59.2% (SD 2%) on the PVI dataset (as presented in Table 3). This is very similar to accuracy obtained using top features identified using only the PVI dataset (mean balanced accuracy of 63.7% (SD 1.8%)). Table 3 provides details of the selected feature subsets identified using the FS algorithms in this study. We remark that there is some similarity in the top selected features from the different FS algorithms, which inspires some confidence in tentatively interpreting findings. Table 4 summarizes the association strength metrics for the most strongly correlated dysphonia measures with the response.

Table 5 summarizes findings in the research literature and this study. The results are presented in the form mean ± standard deviation where appropriate. In Table 5, SVM stands for support vector machine, GP-EM for genetic programming and the expectation-maximization algorithm, E-M for expectation maximization algorithm, and RF-GSO for random forests with Gram-Schmidt orthogonalization scheme. Hitherto, all studies used high-quality data collected where the voice signals were recorded under carefully controlled conditions (e.g. head-mounted microphone, and often IAC booths). This study uses a considerably larger database with data collected under highly non-controlled conditions.

We remark that, until now, studies in the research literature typically use high-quality voice samples that were recorded in an IAC sound-treated booth with a head-mounted microphone; therefore the results are not directly comparable. Nevertheless, we wanted to survey the research literature on the reported accuracy in controlled laboratory settings, since this is the only setting against which the current study's findings could be contrasted. Table 6 summarizes the classification performance results of this study. The best performance using RF and mRMR, and, RF and RELIEF, was obtained using all 307 features, while for RF and GSO, using 100 key identified features gave the best performance (sensitivity = 64.90%±2.90%; specificity = 67.96%±2.90%; balanced accuracy = 66.43%±1.83%). Using

LASSO, employing 200 features resulted in the best performance, while ensemble ranking with 100 features resulted in the highest accuracy. Although the sex ratio for PD and control cohorts are very similar (0.4322 and 0.4481, respectively), to account for any potential differences in sex, we computed sensitivity and specificity separately for all recordings, only female recordings, and only male recordings (as presented in Table 6). Using only female recordings (1199 recordings from 641 PD participants and 6922 recordings from 3719 controls), the highest accuracy was obtained using RF-LASSO with (sensitivity = 65.23%±4.48%; specificity = 63.44%±3.92%; balanced accuracy = 64.34%±2.98%). Similarly, using only male recordings (1560 recordings from 842 PD participants and 8399 recordings from 4581 controls), the highest accuracy was obtained using RF-GSO with (sensitivity = 67.29%±4.01%; specificity = 70.28%±4.12%; balanced accuracy = 68.79%±2.75%). As evident from Table 6, the accuracies obtained using all recordings and sex-specific recordings were quite similar.

The sensitivity, specificity, accuracy and balanced accuracy results differed from comparable results obtained from randomized predictions (denoted as Random Classifier in Table 6) regarding class membership ($p < 0.05$, two-sided Kolmogorov-Smirnov test). Moreover, we also checked for potential bias in the results by randomizing the labels (i.e. dissociating the relationship between the target variable and the corresponding labels). Randomizing the labels resulted in an average classification accuracy of ~50% (SD 2%). Although the expected chance level for a binary classification problem in a balanced dataset is 50%, we felt it was important to validate our findings against naïve benchmarks (such as random classifier, and random forest applied to dataset after randomizing labels), as it has been demonstrated that applying signal classification to Gaussian random signals can result in accuracies of up to 70% or higher in binary class problems with small sample sets (Combrisson and Jerbi, 2015). Moreover, as a benchmark, we used a Naïve Bayes (NB) classifier as it is relatively easy to construct, has low computational costs, and has been shown to be competitive with sophisticated techniques (Domingos and Pazzani, 1997; Kononenko, 1993; Shree and Sheshadri, 2018). Using all recordings, the NB classifier achieved the highest average balanced accuracy of 59.1% (SD 2.3%). Whereas, using female and male recordings with the NB classifier, the average balanced accuracy was 59.7% (SD 3.0%) and 60.9% (SD 4.8%), respectively. For all pairwise comparisons, the discrimination results obtained using RF were considerably better than the corresponding accuracies obtained using randomized predictions and a NB classifier (Table 6).

Finally, following a reviewer's comment we wanted to investigate whether there is any relationship between the most strongly associated features with the clinical outcome (presented in Table 4), to explore whether those features might be useful in assessing presbyphonia. Figure 5 suggests that the dysphonia measures explored in this study could potentially be used to assess presbyphonia, and further work could investigate in further detail the difference between normative values as a function of age across those dysphonia measures and the difference observed in PD.

## V. Discussion

This study investigated the potential of using telephone-quality voice recordings for discriminating PD participants from control participants. It is the largest PD characterization study undertaken using telephone-quality voice recordings, and is a step towards establishing the developed methodology for practical use in screening the population at large for PD. We remark that previous studies in the research literature were considerably more limited in the number of participants; moreover, they relied on high-quality data typically recorded under highly controlled conditions (sound-treated booth, head-mounted microphone, built-for-purpose equipment) and used expert clinical diagnosis as the ground truth.

We report a sensitivity of 64.90%±2.90% and specificity of 67.96%±2.90% in differentiating PD from controls on a balanced dataset. This result is considerably worse compared to studies in the literature that reported almost 98% accuracy in the same application using high-quality data (Tsanas et al., 2012), however, it is crucial to appreciate that the results reported in this study have been obtained: (1) using data collected in a home environment without any supervision or prior training, which results in extraneous background noise and a variety of different user behaviors (such as, distance of phone from the mouth, volume of sustained phonation etc.) ; (2) using a standard telephone network that results in recordings at low sampling frequency (8KHz), which can results in a small number of observations per recording to be used for feature extraction; (3) from participants who self-reported their symptoms (as either PD/HC), and thus we cannot rule out the presence of a variety of clinical-pathologic differences in voice within this cohort. We remark that most previous studies have typically collected data in a laboratory-environment using high-quality microphones in sound-treated booths that minimize background noise, whereby the clinical assessments are done by experts. Despite the simplicity of our experimental design, we emphasize that our findings may have a considerable practical impact in resource-constrained settings and for readily available, cost-effective screening of the population for PD when lacking specialized facilities.

Compared to other studies that have looked into the same problem, we have found considerably lower correlations between the features and the binary response. For example,

Tsanas et al. (2012) had reported that some features exhibited correlation coefficient magnitudes that were over 0.3, i.e. correlations that may be considered statistically strong in the medical domain (Tsanas et al., 2013). Therefore, these exploratory analysis findings already suggested that the classification performance in this study would likely be worse compared to previous studies.

We have used four robust, widely studied FS algorithms to identify feature subsets (Table 3). Overall, we note that the algorithms are broadly in agreement towards the selected features (or feature families). We remark that features which are related to F0 and frequency variability tend to dominate. This is not surprising, since participants may have been holding their phone's microphone at different distances from their mouth, which would have affected the recorded amplitude (therefore, dysphonia measures quantifying frequency aspects that can be considered more robust). Similarly, some of the MFCCs were selected, in accordance to previous studies that have reported that MFCCs empirically work well in biomedical signal processing applications (Godino-Llorente et al., 2006; Tsanas et al., 2011). It is difficult to interpret the physical meaning of MFCCs: broadly, lower MFCCs quantify the amplitude and spectral envelope and higher MFCCs quantify information about harmonic components. Interestingly, some of the nonlinear measures which have previously worked very well under the controlled setting setups have not been selected amongst the top choices of the FS algorithms. We attribute this to the fact that the nonlinear dysphonia measures rely on highly sensitive characteristics of the speech signals. Further work is warranted to verify the present study's findings and determine whether the nonlinear dysphonia measures suffer in settings where we lack high-quality voice signals, and in cases where we have bandlimited and potentially noisy recording environments.

We remark that in this study we have focused exclusively on the sustained vowel /a/ and the extraction of *dysphonia measures*. Dysarthria is a key characteristic in PD and can be assessed using conversational speech. It is possible that MFCC might be capturing some of the dysarthric components, which might explain, at least in part, their success in similar biomedical applications. Future work could look into whether the information extracted from sustained vowels using MFCCs is associated with dysarthria symptoms that can be extracted from conversational speech.

The statistical mapping used in this study falls under the standard supervised learning umbrella with a binary classification setting. As such, there is a multitude of available statistical machine learning algorithms (e.g. see Hastie et al., (2009) for an overview) that aim to identify a functional mapping of the feature set to the response (in this study, the binary outcome of PD vs controls). Future studies could experiment using SVMs, Gaussian processes, boosting approaches (Adaboost and other robust boosting approaches) and compare these results reported using RF. It would be potentially interesting to also experiment using the probabilistic outcomes of the various classifiers, and use their outputs as features in the *second* layer of classifiers. The findings of this study warrant further investigations to better understand the effect of noise and low sampling rate on voice for distinguishing PD and controls. Moreover, future studies could investigate automated voice segmentation and noise removal algorithms for preprocessing the voice recordings collected under non-clinical settings.

We envisage this study as a step towards the larger goal of technologies for developing diagnostic decision support tools in PD. Furthermore, we remark that the healthy subjects in previous studies did not have any pathological vocal symptoms when assessed by expert speech scientists (e.g. see Tsanas et al., 2012). One of the major limitations of this study is that it did not include individuals with other Parkinsonian or voice-related disorders that may be more difficult to differentiate from PD. It is, however, possible that this study might include a cohort of subjects with potential PD-like vocal symptoms, who do not qualify for PD diagnosis otherwise. The proposed methodology cannot be readily validated or used in a general population as a screening tool for PD, as there are different types of dysphonia against which validation needs to be performed. It is also worth highlighting that the vocal pathologies associated with PD are complex; differences have been reported recently in voice impairment between participants with idiopathic PD and leucine-rich repeat kinase 2 (*LRRK2*)-associated PD (Arora et al., 2018b), while there is also evidence of speech impairment in participants who are at risk of developing PD, i.e. participants with RBD (Rusz et al., 2016). *LRRK2* is the greatest known genetic factor associated with PD (Healy et al., 2008), while RBD is the strongest known predictor for PD (Iranzo et al., 2013). It can be envisaged that the PD participants who took part in this study could either be idiopathic or *LRRK2*-associated, while this study might also include a cohort of control participants with potential PD-like vocal symptoms (including idiopathic RBD), who do not qualify for PD diagnosis otherwise. Hence, although this study did not explicitly focus on a broad range of vocal pathologies associated with other Parkinsonian or voice-related disorders, one cannot rule out the presence of PD and control participants who exhibit a variety of clinical-pathologic differences in voice within this cohort. We remark that this study only looked into the problem of binary differentiating PD from HC, which builds upon previous work on voice impairment in PD (see Arora et al., 2015; Åström and Koker, 2011; Guo et al., 2010; Little et al., 2009; Orozco-Arroyave et al,. 2016; Tsanas et al., 2011; Tsanas et al., 2012; Tsanas et al., 2014b). It would be interesting to use a very large database including voices from diverse disorders, where the use of sophisticated dysphonia measures might assist in determining the underlying pathology amongst a wide set of possible diagnoses.

## Acknowledgments

We are grateful to M.A. Little who led the PVI where the data for this study was collected. We would like to extend our sincere gratitude to all the participants who took part in the PVI study. The study was made possible through generous funding via an EPSRC-NCSML award to SA and AT.

## Conflict of interest

None to declare.

**TABLE 1**
BASELINE CHARACTERISTICS OF ALL STUDY PARTICIPANTS

| Characteristics | Parkinson's Disease Participants | Control Participants |
|---|---|---|
| **A.  All participants** | | |
| # of participants | 1483 | 8300 |
| Mean age (standard dev.) | 62.87 years (11.41 years) | 47.74 years (15.69 years) |
| % Female | 0.4322 | 0.4481 |
| # of voice rec./participant | 1.86 | 1.85 |
| # total usable recordings | 2759 | 15321 |
| **B.  Female participants** | | |
| # of participants | 641 | 3719 |
| Mean age (standard dev.) | 62.05 years (11.61 years) | 49.90 years (15.02 years) |
| % Female | 1 | 1 |
| # of voice rec./participant | 1.87 | 1.86 |
| # total usable recordings | 1199 | 6922 |
| **C.  Male participants** | | |
| # of participants | 842 | 4581 |
| Mean age (standard dev.) | 63.49 years (11.22 years) | 45.98 years (16.01 years) |
| % Female | 0 | 0 |
| # of voice rec./participant | 1.85 | 1.83 |
| # total usable recordings | 1560 | 8399 |

**TABLE 2**
BREAKDOWN OF THE 307 DYSPHONIA MEASURES USED IN THIS STUDY

| Family of dysphonia measures | Brief description | Number of measures |
|---|---|---|
| Jitter variants | F0 perturbation | 28 |
| Shimmer variants | Amplitude perturbation | 21 |
| Harmonics to noise ratio (HNR) and noise to harmonics ratio (NHR) | Signal to noise, and noise to signal ratios | 4 |
| Glottis quotient (GQ) | Vocal fold cycle duration changes | 3 |
| Recurrence period density entropy (RPDE) | Uncertainty in estimation of fundamental frequency | 1 |
| Detrended fluctuation analysis (DFA) | Stochastic self-similarity of turbulent noise | 1 |
| Pitch period entropy (PPE) | Inefficiency of F0 control | 1 |
| Glottal to noise excitation (GNE) | Extent of noise in speech using energy and nonlinear energy concepts | 6 |
| Vocal fold excitation ratio (VFER) | Extent of noise in speech using energy, nonlinear energy, and entropy concepts | 9 |
| Empirical mode decomposition excitation ratio (EMD-ER) | Signal to noise ratios using EMD-based energy, nonlinear energy and entropy | 6 |
| Mel Frequency Cepstral Coefficients (MFCC) | Amplitude and spectral fluctuations | 42 |
| F0-related measures | Summary statistics of F0, Differences from expected F0 in age- and sex- matched controls, variations in F0 | 3 |
| Wavelet decomposition measures | Decomposition of the F0 contour to derive transient characteristics | 182 |

Algorithmic expressions for the 307 measures summarized here are described in detail in Tsanas (2012). F0 refers to fundamental frequency estimates.

| mRMR | GSO | RELIEF | LASSO | ENSEMBLE RANKING | TSANAS ET AL 2012 |
|---|---|---|---|---|---|
| mean HNR {17} | 6th delta MFCC {12} | HNR {17} | Jitter (TKEO 25% pitch) {4} | HNR {17} | 10th detailed wavelet coef. std TKEO {176} |
| Log energy (MFCC) {49} | 6th detailed wavelet coef. log entropy of $F_0$ {1}the F0 | Log energy (MFCC) {49} | Jitter (TKEO 95% pitch) {6} | Log energy (MFCC) {49} | 4th detailed wavelet coef. std TKEO {170} |
| Jitter (TKEO of 25% $F_0$) {5} | Standard deviation of the TKEO of 1st approximate wavelet coef. {168} | Jitter (TKEO 95% pitch) {7} | Jitter mean TKEO pitch {222} | Jitter (TKEO 95% pitch) {6} | $VFER_{SNR\text{-}TKEO}$ {73} |
| Jitter (TKEO 95% pitch) {6} | Standard deviation of the TKEO of 2nd approximate wavelet coef. {166} | Jitter (TKEO 25% pitch) {4} | Jitter (TKEO 5% pitch) {269} | Jitter (TKEO 25% pitch) {4} | HNR {17} |
| Jitter (TKEO of 95% $F_0$) {7} | 0th MFCC coef {28} | 6th detailed wavelet coef. log entropy of $F_0$ {2} | Jitter std. TKEO pitch {285} | Jitter (TKEO of 25% $F_0$) {5} | $VFER_{SNR\text{-}TKEO}$ {71} |
| Jitter (TKEO 25% pitch) {4} | Jitter (TKEO 95% pitch) {7} | 6th detailed wavelet coef. log entropy of $F_0$ {1} | Jitter absolute pitch {146} | 6th detailed wavelet coef. log entropy of $F_0$ {1} | GNE std. {61} |
| 6th detailed wavelet coef. log entropy of $F_0$ {1} | 4th MFCC coef. {27} | Jitter (TKEO of 95% $F_0$) {7} | Jitter pitch PQ5 {304} | Jitter (TKEO of 95% $F_0$) {7} | 12th MFCC coef. {94} |
| 6th detailed wavelet coef. Energy of log $F_0$ {177} | 1st delta MCC coef. {13} | 7th detailed wavelet coef. log ($F_0$) entropy {8} | 1st detailed wavelet coef. Energy of log $F_0$ {304} | Jitter absolute pitch {146} | 6th detailed wavelet coef. mean TKEO {162} |
| 6th detailed wavelet coef. Energy of $F_0$ {261} | 9th MFCC coef. {40} | 7th detailed wavelet coef. ($F_0$) entropy {3} | 10th delta-delta MFCC {109} | Jitter pitch PQ5 {304} | 11th MFCC coef. {32} |
| Mean TKEO of 6th detailed wavelet coef. {201} | Difference $F_0$ and age- & gender-matched $F_0$ {18} | Jitter (TKEO of 25% $F_0$) {5} | 9th delta-delta MFCC {114} | Jitter mean TKEO pitch {222} | Jitter pitch PQ5 {304} |
| 60.1%±1.9% | 63.7%±1.8% | 59.4%±2.0% | 60.1%±2.1% | 60.9%±2.1% | 59.2%±2.0% |

The last row presents the % balanced accuracy (computed as the mean of specificity and sensitivity) when the top ten selected features from each algorithm are fed into the classifier. The results are given in the form mean ± standard deviation and are out of sample computed using 10-fold cross validation with 10 repetitions using a balanced dataset. The number in curly brackets '{}' indicates the rank of the correlation of the feature with the binary outcome (PD vs HC) when computing the correlation coefficient for all 307 features, e.g. {2} would suggest that this was the second most correlated feature with the outcome.

**TABLE 4**

STATISTICAL ANALYSIS OF THE DYSPHONIA FEATURES

| Dysphonia measure | Correlation coefficient |
|---|---|
| Entropy of the 6[th] detail wavelet decomposition coefficient of the F0 | 0.11±0.00 |
| Entropy of the 6[th] detail wavelet decomposition coefficient of the log-transformed F0 | 0.10±0.00 |
| Entropy of the 7[th] detail wavelet decomposition coefficient of the F0 | 0.10±0.00 |
| Jitter TKEO 25[th] percentile of pitch | -0.09±0.00 |
| Jitter TKEO 25[th] percentile of F0 | -0.09±0.00 |
| Jitter TKEO 95[th] percentile of pitch | 0.09±0.00 |
| Jitter TKEO 95[th] percentile of F0 | 0.08±0.00 |
| Entropy of the 7[th] detail wavelet decomposition coefficient of the log-transformed F0 | 0.08±0.00 |
| Entropy of the 5[th] detail wavelet decomposition coefficient of the F0 | 0.08±0.00 |
| Entropy of the 4[th] detail wavelet decomposition coefficient of the F0 | 0.08±0.00 |

Ten features most strongly associated with the response, sorted using the magnitude of the correlation coefficient. All reported correlations were statistically significant ($p < 0.001$). We used a jack-knife approach by randomly sampling the over-populated class to obtain balanced datasets and compute the correlations; the results are in the form mean±std from 100 repetitions. Also, the results of the Mann Whitney statistical test suggest all relationships are statistically significant ($p < 0.001$). The response was '0' for healthy controls and '1' for people with Parkinson's disease. Thus, positive correlation coefficients suggest that the dysphonia measure takes, in general, larger values for Parkinson's disease phonations.

**TABLE 5**

| Study | Study population and brief details | Learning and validation scheme | Reported accuracy (%) |
|---|---|---|---|
| Little et al., 2009 | 31 participants (23 PD), 195 phonations recorded in an IAC sound-treated booth using a head-mounted microphone. Use of sustained /a/ vowels | Support vector machine (SVM), bootstrap with 50 replicates | 91.4 ± 4.4 |
| Guo et al., 2010 | 31 participants (23 PD) from the study of Little et al. (2009). | Genetic programming and expectation maximization (GP-EM), 10-fold cross-validation | 93.1 ± 2.9 |
| Das, 2010 | 31 participants (23 PD) from the study of Little et al. (2009). | Neural Network, 65% data for training, rest for testing | 92.9 |
| Åström and Koker, 2011 | 31 participants (23 PD) from the study of Little et al. (2009). | Neural Network, 60% data for training, rest for testing | 91.2 |
| Tsanas et al., 2012 | 43 subjects (33 PD), 263 phonations, recorded in an IAC sound-treated booth using a head-mounted microphone. Use of sustained /a/ vowels | SVM, 10-fold cross-validation | 97.7 ± 2.8 |
| Chen et al., 2013 | 31 participants (23 PD) from the study of Little et al. (2009). | Principal component analysis and fuzzy k-nearest neighbour based model, 10-fold cross-validation | 96.07 |
| Orozco-Arroyave et al,. 2016 | Data from Spanish, German, and Czech cohorts. 100 native Spanish speakers (50 PD), 176 German speakers (88 PD), 36 Czech speakers (20 PD). Use of continuous speech, and pa/ta/ka tests | SVM, 10-fold cross-validation, and leave one speaker out | 85 to 99 |
| This study* | 9783 subjects (1483 PD), 18080 phonations collected under non-controlled conditions using the standard telephone system. Use of sustained /a/ vowels | RF-GSO, 10-fold cross-validation | 66.4 ± 1.8 |

CLASSIFICATION ACCURACY USING SELECTED FEATURE SUBSETS AND DIFFERENT CLASSIFICATION
SCHEMES USING ALL RECORDINGS

| CLASSIFIER + FEATURE SELECTION METHOD | SENSITIVITY | SPECIFICITY | ACCURACY | BALANCED ACCURACY |
|---|---|---|---|---|
| **A.   ALL RECORDINGS** | | | | |
| RF-mRMR (307 Features) | 63.9 %±3.0% | 67.4 %±3.0% | 65.7 %±2.2% | 65.7%±2.2% |
| RF-GSO (100 Features) | **64.9%±2.9%** | **68.0%±2.9%** | **66.4%±1.8%** | **66.4%±1.8%** |
| RF-RELIEF (307 Features) | 63.9%±3.0% | 67.4%±3.0% | 65.6%±2.2% | 65.6%±2.2% |
| RF-LASSO (200 Features) | 64.1%±2.9% | 68.0%±2.4% | 66.1%±1.8% | 66.1%±1.8% |
| RF-Ensemble Ranking (100 Features) | 64.1%±2.6% | 68.3%±2.2% | 66.2%±1.6% | 66.2%±1.6% |
| Naïve Bayes-GSO | 60.9%±3.8% | 57.2%±3.7% | 59.1%±2.2% | 59.1%±2.3% |
| **B.   FEMALE RECORDINGS** | | | | |
| RF-mRMR (307 Features) | 65.1%±4.8% | 63.4 %±4.2% | 64.2 %±3.3% | 64.2%±3.3% |
| RF-GSO (150 Features) | **64.9%±4.4%** | **63.7%±4.7%** | **64.3%±3.1%** | **64.3%±3.2%** |
| RF-RELIEF (307 Features) | 65.1%±4.8% | 63.4%±4.2% | 64.2%±3.3% | 64.2%±3.3% |
| RF-LASSO (150 Features) | **65.2%±4.5%** | **63.4%±3.9%** | **64.3%±2.9%** | **64.3%±3.0%** |
| RF-Ensemble Ranking (307 Features) | 65.1%±4.8% | 63.4%±4.2% | 64.2%±3.3% | 64.2%±3.3% |
| Naïve Bayes-GSO | 52.8%±4.8% | 66.5%±3.9% | 59.6%±3.1% | 59.7%±3.0% |
| **C.   MALE RECORDINGS** | | | | |
| RF-mRMR (307 Features) | 64.4%±4.0% | 69.8 %±3.7% | 67.1 %±2.6% | 67.1%±2.6% |
| RF-GSO (25 Features) | **67.3%±4.0%** | **70.3%±4.1%** | **68.8%±2.8%** | **68.8%±2.8%** |
| RF-RELIEF (150 Features) | 64.2%±3.8% | 70.1%±3.6% | 67.1%±2.5% | 67.1%±2.5% |
| RF-LASSO (150 Features) | 65.8%±3.7% | 69.0%±3.7% | 67.4%±2.5% | 67.4%±2.5% |
| RF-Ensemble Ranking (150 Features) | 65.1%±3.8% | 70.4%±3.8% | 67.7%±2.7% | 67.8%±2.7% |
| Naïve Bayes-GSO | 47.5%±16.8% | 74.4%±8.8% | 60.4%±5.3% | 60.9%±4.8% |

The results are presented in the form mean ± standard deviation where appropriate. For each feature selection scheme, the number of key features that give the best classification results (as quantified using the balanced accuracy) are reported in brackets. Model performance (in %) was quantified using sensitivity = TP/(TP+FN), specificity = TN/(TN+FP), accuracy = (TP+TN)/(TP+FP+FN+TN), and balanced accuracy = (sensitivity + specificity)/2, where TP denotes true positive, TN denotes true negative, FP denotes false positive, and FN denotes false negative. For a given performance measure, the highest classification accuracy is highlighted in **bold**.

**Figure 1**. Schematic diagram illustrating the procedure for collecting voice recordings (sustained vowel 'aaah'), using a standard telephone network, along with the major steps involved in the statistical data analysis. **Step 1** involves collecting voice samples from controls and participants with PD. The raw voice recordings are pre-processed in order to identify the participant prompts and discard non-usable recordings (unclear prompts or insufficient phonation length). **Step 2** involves extracting features (or summary measures) that quantify key properties of voice such as: reduced loudness, breathiness, roughness, monopitch, and exaggerated vocal tremor, which are commonly associated with voice impairment in PD. **Step 3** identifies the key features, using four feature selection techniques. **Step 4** involves mapping the key identified features onto a clinical assessment (PD/Control). The out-of-sample classification accuracy is measured using a 10-fold cross-validation scheme on a balanced dataset. Abbreviations: DFA = detrended fluctuation analysis; $F_0$ = fundamental frequency; PD = Parkinson's disease; RPDE = recurrence period density entropy.

**Figure 2**. Out-of-sample comparison of the feature selection algorithms obtained using *all recordings* (2759 recordings from 1483 PD participants, and 15321 recordings from 8300 control participants), based on learner performance (binary-class classification datasets) using Random Forests (RF). The classification accuracy is computed on a balanced-dataset using 10-fold cross-validation scheme with 10 repetitions. The classification accuracy is quantified using sensitivity (%) and specificity (%), whereby sensitivity = TP/(TP+FN) and specificity = TN/(TN+FP), where TP denotes true positive, TN denotes true negative, FP denotes false positive, and FN denotes false negative. The horizontal axis denotes the different number of features (10, 25, 50, 75, 100, 150, 200, 250, 307) selected across all four feature selection algorithms and the ensemble ranking scheme.

**Figure 3**. Out-of-sample comparison of the feature selection algorithms obtained using *only female recordings* (1199 recordings from 641 PD participants, and 6922 recordings from 3719 control participants), based on learner performance (binary-class classification datasets) using Random Forests (RF). The classification accuracy is computed on a balanced-dataset using 10-fold cross-validation scheme with 10 repetitions. The classification accuracy is quantified using sensitivity (%) and specificity (%). The horizontal axis denotes the different number of features (10, 25, 50, 75, 100, 150, 200, 250, 307) selected across all four feature selection algorithms and the ensemble ranking scheme

**Figure 4**. Out-of-sample comparison of the feature selection algorithms obtained using *only male recordings* (1560 recordings from 842 PD participants, and 8399 recordings from 4581 control participants), based on learner performance (binary-class classification datasets) using Random Forests (RF). The classification accuracy is computed on a balanced-dataset using 10-fold cross-validation scheme with 10 repetitions. The classification accuracy is quantified using sensitivity (%) and specificity (%). The horizontal axis denotes the different number of features (10, 25, 50, 75, 100, 150, 200, 250, 307) selected across all four feature selection algorithms and the ensemble ranking scheme

**Figure 5**. Scatter plots to visually assess the relationship between the most strongly associated dysphonia measures with clinical outcomes (summarized in Table 4), to explore whether those could be used to assess presbyphonia.

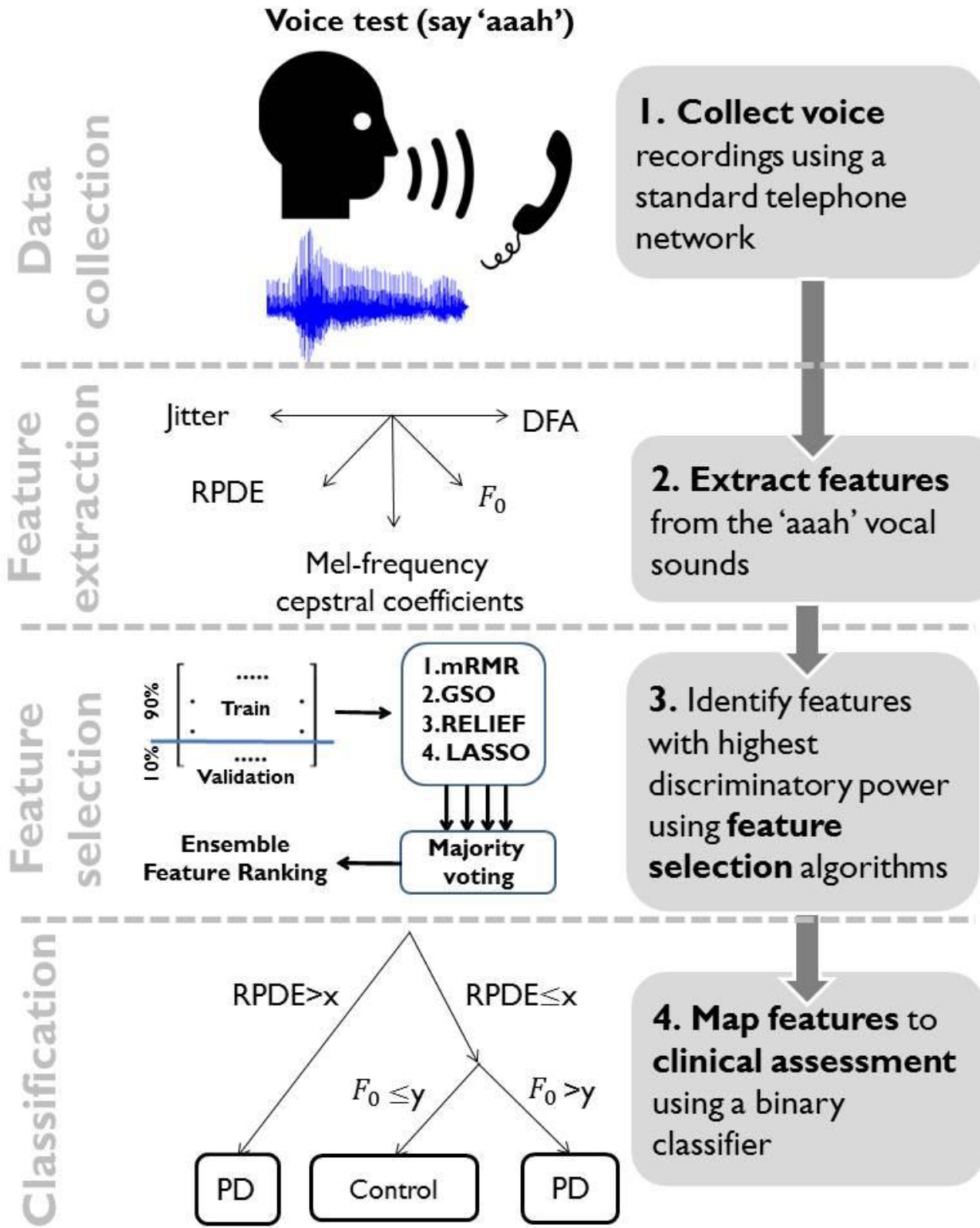

FIGURE 2.

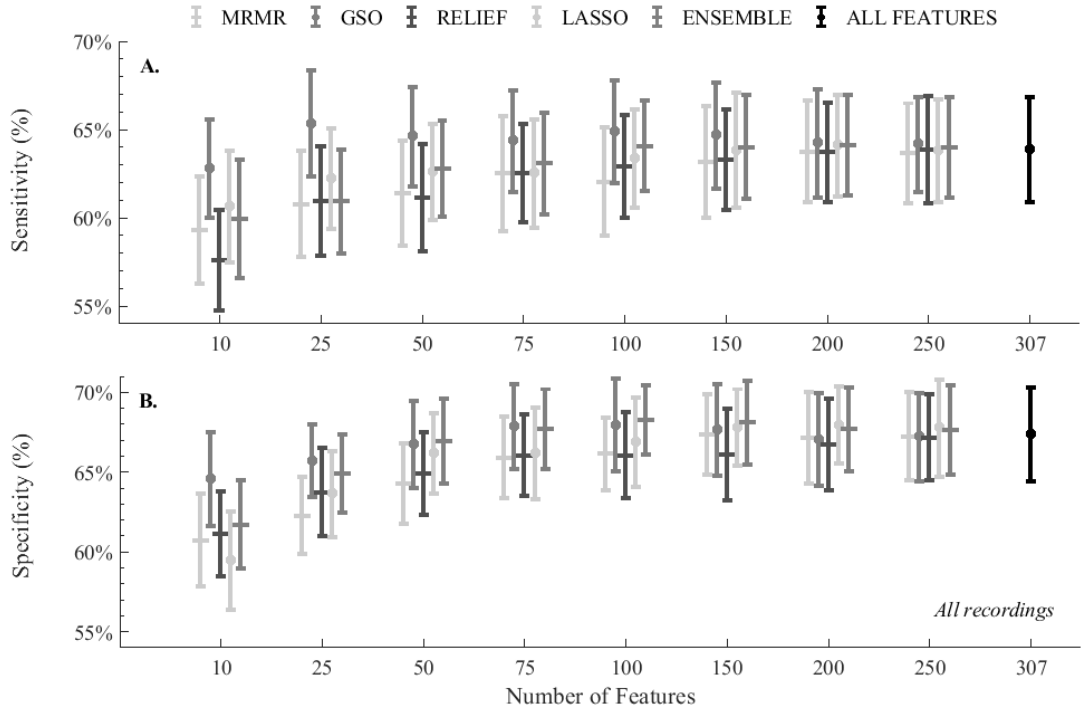

FIGURE 3.

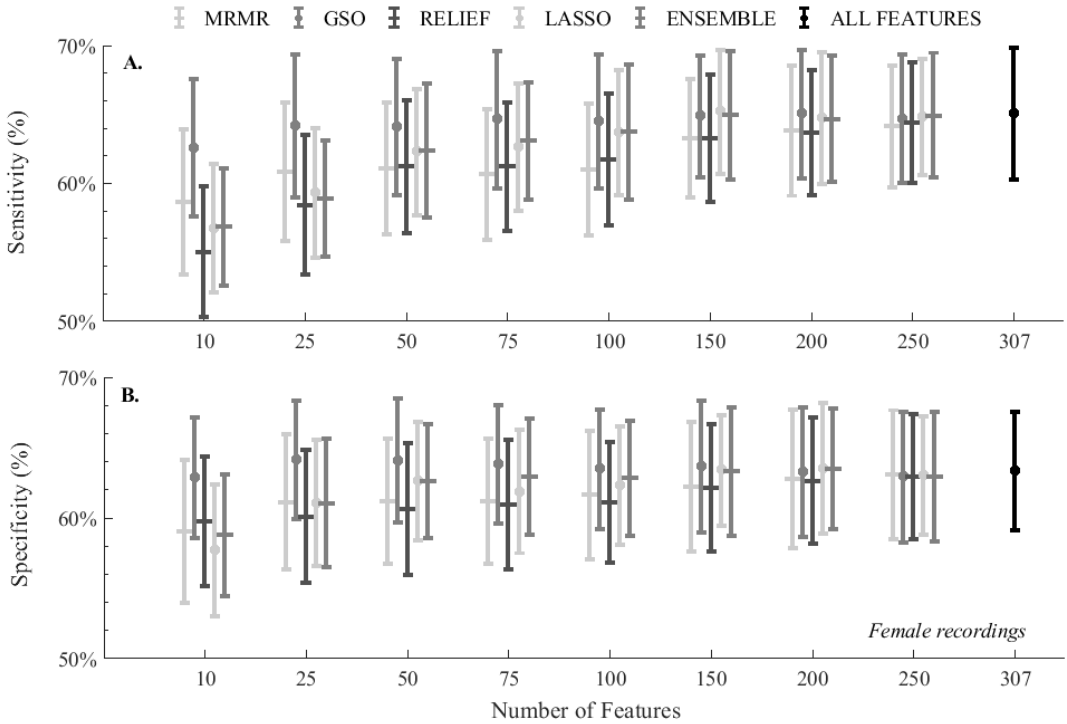

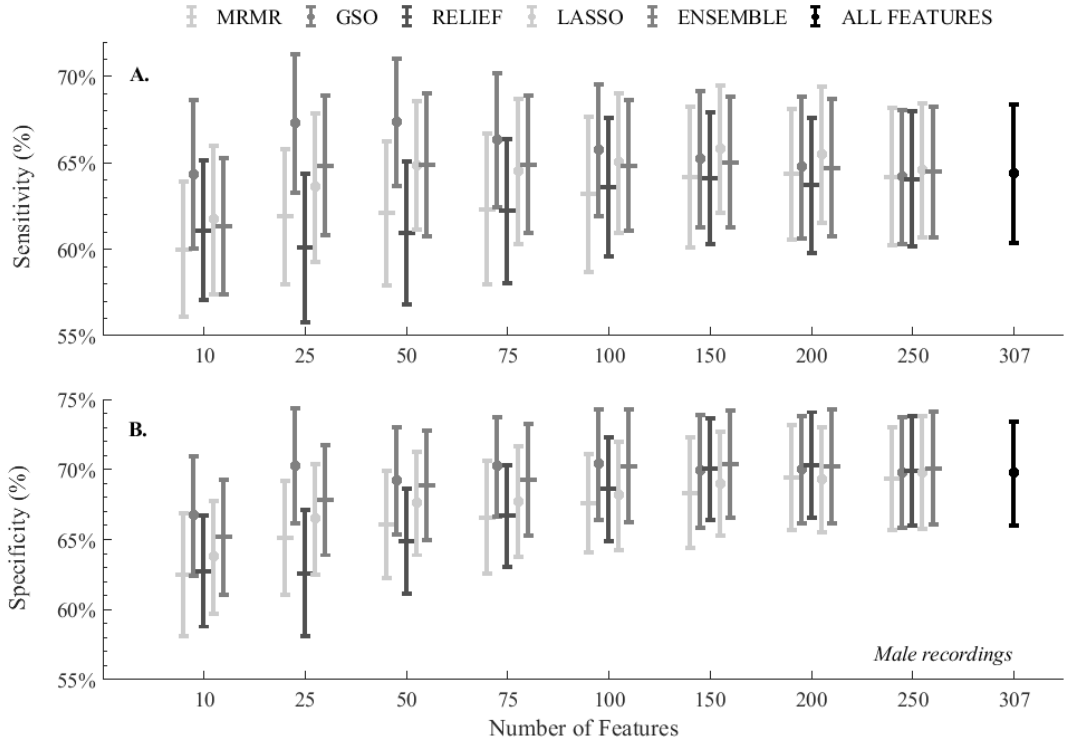

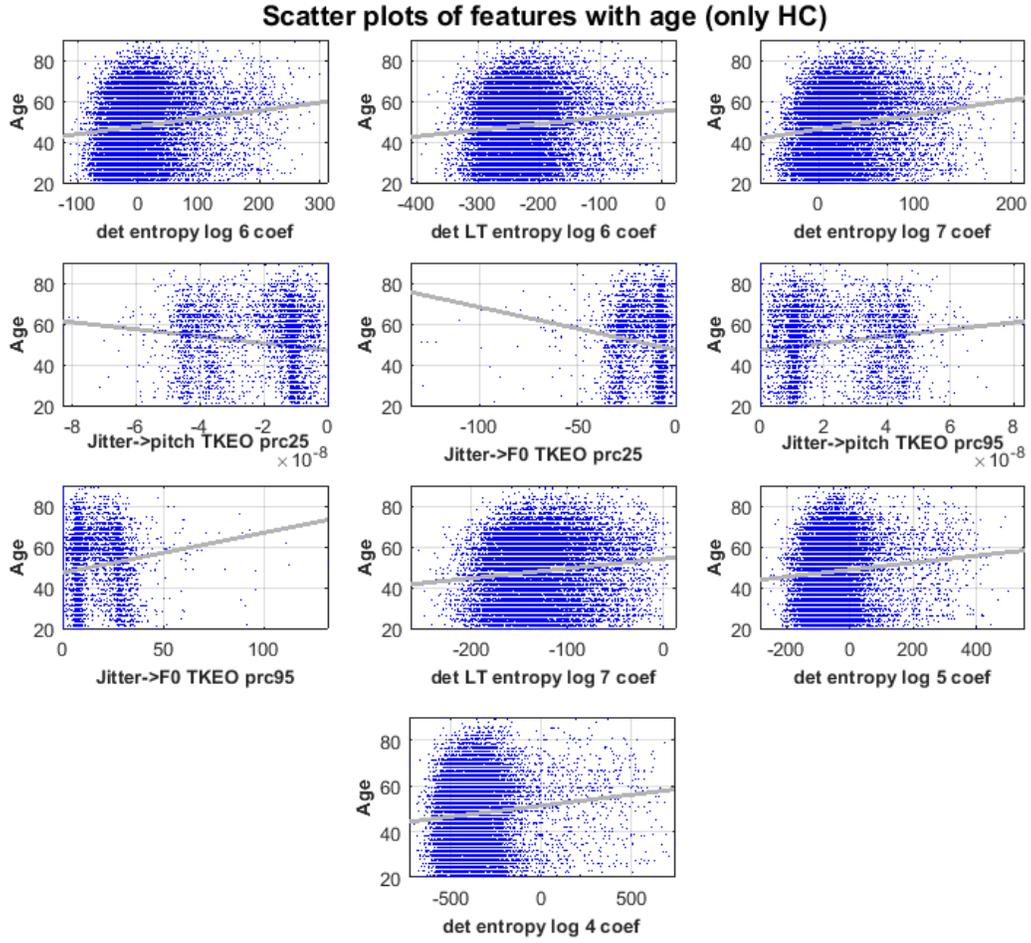